\documentclass[12pt,english]{article}
\usepackage{ae,aecompl}
\usepackage[T1]{fontenc}
\usepackage[latin1]{inputenc}
\usepackage{geometry}
\usepackage{rotating}
\usepackage{booktabs}
\usepackage{amsmath}
\usepackage{color}
\usepackage{graphicx}
\usepackage{setspace}
\usepackage{amssymb}

\makeatletter

\providecommand{\tabularnewline}{\\}

\newcommand{\lyxaddress}[1]{
\par {\raggedright #1
\vspace{1.4em}
\noindent\par}
}

\usepackage[super]{cite}

\usepackage{babel}
\makeatother

\begin{document}

\title{\textbf{\large Scaling factors for ab initio vibrational frequencies:}\\
\textbf{\large comparison of uncertainty models for quantified prediction}}

\author{Pascal Pernot$^{1,2}$}

\maketitle

\lyxaddress{$^{1}$Laboratoire de Chimie Physique, CNRS, UMR8000, Orsay, F-91405\\
$^{2}$Univ Paris-Sud, Orsay, F-91405}

\begin{abstract}
\noindent Bayesian Model Calibration is used to revisit the problem
of scaling factor calibration for semi-empirical correction of \emph{ab
initio} calculations. A particular attention is devoted to uncertainty
evaluation for scaling factors, and to their effect on prediction
of observables involving scaled properties. We argue that linear models
used for calibration of scaling factors are generally not statistically
valid, in the sense that they are not able to fit calibration data
within their uncertainty limits. Uncertainty evaluation and uncertainty
propagation by statistical methods from such invalid models are doomed
to failure. To relieve this problem, a stochastic function is included
in the model to account for model inadequacy, according to the Bayesian
Model Calibration approach. In this framework, we demonstrate that
standard calibration summary statistics, as optimal scaling factor
and root mean square, can be safely used for uncertainty propagation
only when large calibration sets of precise data are used. For small
datasets containing a few dozens of data, a more accurate formula
is provided which involves scaling factor calibration uncertainty.
For measurement uncertainties larger than model inadequacy, the problem
can be reduced to a weighted least squares analysis. For intermediate
cases, no analytical estimators were found, and numerical Bayesian
estimation of parameters has to be used.\emph{}
\end{abstract}
\newpage{}

\section{Introduction}

The final stage in the development of a model, after calibration on
an experimental dataset and proper validation, consists in its use
for prediction: ''If the experimental dataset is sufficiently broad,
there is a reasonable expectation that the results will be accurate
to something like the target accuracy''.\cite{Pople99} The estimation
of uncertainty for the results of computational chemistry is indeed
of paramount importance, notably to their use in multiscale modeling.
\cite{Pancerella03,Frenklach07} The concept of Virtual Measurement
has been introduced by Irikura\cite{Irikura04} to take advantage
of the standardized procedures defined by the Guide to the Expression
of Uncertainty in Measurement (also known as ''the GUM''),\cite{GUM}
and to apply them in the case of quantum chemistry modeling. Random
uncertainties have been shown to be very small, and the major uncertainty
factor are biases due to basis-set and/or theory limitations. For
quantum chemistry to be predictive, i.e. to be able to predict observables
with confidence intervals, one has to correct for these biases, which
commonly requires semi-empirical corrections. Uncertainty attached
to these corrections have to be considered in the final uncertainty
budget, to which they constitute often a major contribution.

\begin{figure}[p]
\noindent \begin{centering}
\includegraphics[bb=0bp 0bp 612bp 612bp,clip,width=0.75\columnwidth]{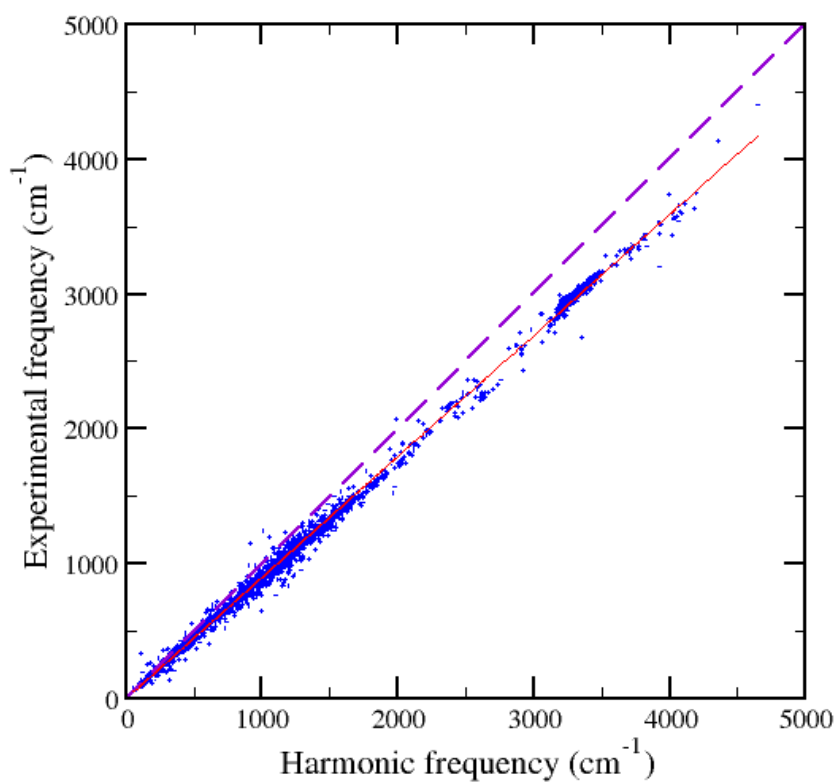}
\par\end{centering}

\caption{\label{fig:Correlation-plot}Correlation plot between calculated harmonic
frequencies $\omega_{i}$ and measured frequencies $\nu_{i}$ for
a set of vibrations extracted from the CCCBDB for the HF/6-31G{*}
combination of theory/basis-set (dots). The full line is the regression
line $\nu=s\omega$; the dashed line is a visual aid to appreciate
the bias.}

\end{figure}

Scaling of harmonic vibrational frequencies is an important example
of calibration method in computational chemistry, where estimation
of a vibrational frequency $\nu$ is obtained by multiplying the corresponding
harmonic vibrational frequency $\omega$, routinely calculated by
standard computational chemistry codes, by an empirical scaling factor
$s$ (Fig. \ref{fig:Correlation-plot})\begin{equation}
\nu=\omega\, s\end{equation}
Optimal scaling factors have been computed for extensive sets of theory/basis-set
combinations \cite{Scott96,Wong96,Merrick07,cccbdb}. In the majority
of publications about scaling factors, two summary statistics are
provided for each theory/basis-set combination: the optimal scaling
factor and the root mean squares. From a reference dataset of experimental
$\left\{ \nu_{i}\right\} _{1}^{N}$ and calculated vibrational frequencies
$\left\{ \omega_{i}\right\} _{1}^{N}$ the optimal scaling factor
obtained by the least-squares procedure is\begin{equation}
\hat{s}=\sum\omega_{i}\nu_{i}/\sum\omega_{i}^{2}\label{eq:optim_s}\end{equation}
and the quality of the correction is estimated by the root mean squares
(rms) value \begin{align}
\gamma & =\left(\frac{1}{N}\sum\left(\nu_{i}-\hat{s}\omega_{i}\right)^{2}\right)^{1/2}\label{eq:rms}\end{align}
To our knowledge, these values are not explicitly used for uncertainty
propagation, but the rms provides an estimate of the residual uncertainty
resulting from the scaling correction (''something like the target
accuracy''\cite{Pople99}, or ''a surrogate for uncertainty'' according
to \cite{Irikura05}).

In a recent article (hereafter referred to as Paper I), Irikura \emph{et
al.}\cite{Irikura05} treated the problem of uncertainty propagation
from scaled frequencies, and, using procedures advocated by the GUM
\cite{GUM}, insisted that the scaling factor itself is subject to
uncertainty, and proposed that this uncertainty is the major contribution
to prediction uncertainty. On the practical side, prediction uncertainty
would be proportional to the calculated harmonic frequency; on the
factual side, these authors declare that scaling factors for vibrational
frequencies are accurate to only two significant figures, whereas
all other studies overstate their precision by reporting them with
four figures. This approach has been adopted by the National Institute
of Standards and Technology (NIST) and put into practice in the Computational
Chemistry Comparison and Benchmark DataBase (CCCBDB) \cite{cccbdb},
section XIII.C.2, where all scaling factors are provided with uncertainties
derived according to the procedure of Paper I\emph{.}

In the present paper, we revisit the problem of scaling factor calibration
and we recast it in the Bayesian Model Calibration framework, reputed
for providing consistent uncertainty evaluation and propagation \cite{Kennedy01,Gregory05,GUMSupp1}.
Section \ref{sec:Methods} presents the methodological elements used
for calibration and validation procedures, which are applied to representative
vibrational frequency and zero point vibrational energy datasets in
Section \ref{sec:Applications-and-discussion}. Bayesian calculations
used in this study are fairly standard, but for the sake of completeness
and for readers unfamiliar with statistical modeling, details are
provided in the Appendix .

\section{Methods\label{sec:Methods}}

\subsection{Statistical model for scaling factor calibration}

Considering a measured frequency $\nu^{obs}$, one can assume that
it is related to the true value $\nu^{true}$ by \begin{equation}
\nu^{obs}=\nu^{true}+\epsilon\end{equation}
where $\epsilon\sim N(0,\rho^{2})$ is a normal random variable centered
at zero, with variance $\rho^{2}$, representing the measurement uncertainty
in the hypothesis of additive white noise. In the following, we assume
that $\rho$ is known beforehand.

Calculated harmonic vibrational frequencies $\omega$ are affected
by random errors, related to numerical convergence defined by non-zero
thresholds and the choice of starting point in geometry optimization,
and to non-zero thresholds in wave-function optimization \cite{Irikura04}.
It has been shown that the associated uncertainties are negligible
when compared to the measurement uncertainties \cite{Irikura04}.
In the following, one can thus assume that the harmonic vibrational
frequencies, being deterministically calculated, have fixed values.

If one makes the hypothesis of a linear relationship between $\nu^{true}$
and $\omega$, the calibration model resulting from this analysis
is \begin{equation}
\nu^{obs}=s\omega+\epsilon\label{eq:simple model}\end{equation}
For a single frequency, there is an optimal scaling parameter $\hat{s}=\nu^{obs}/\omega$.
As $\nu^{obs}$ is uncertain, with standard uncertainty $\rho$, the
value of $s$ cannot be known exactly and has a standard uncertainty
$u_{s}=\rho/\omega$. For a calibration dataset with uniform measurement
uncertainty $\rho$, it can be shown that the optimal value for $s$
is still given by Eq. \ref{eq:optim_s}, and its standard uncertainty
by $u_{s}=\rho/\sqrt{\sum_{i=1}^{N}\omega_{i}^{2}}$ (\emph{c.f.}
Appendix \ref{sub:Case-of-very}). Applicability of this formula is
subject to one condition: the model (Eq. \ref{eq:simple model}) should
be statistically valid. This means that the residuals $\left\{ \nu_{i}^{obs}-\hat{s}\omega_{i}\right\} $
should have a normal distribution with variance $\rho^{2}$. Normality
is not always verified, but most important, the variance condition
is typically violated when precise data are used for calibration.
The linear model (Eq. \ref{eq:simple model}) is typically unable
to reproduce a given set of measured frequencies \emph{within their
uncertainty range}. In such a case, the width of the distribution
of residuals is dominated by model misfit, not by measurement uncertainty,
and the model is invalid.

\begin{figure}[p]
\noindent \begin{centering}
\includegraphics[bb=20bp 30bp 520bp 530bp,clip,width=0.5\columnwidth]{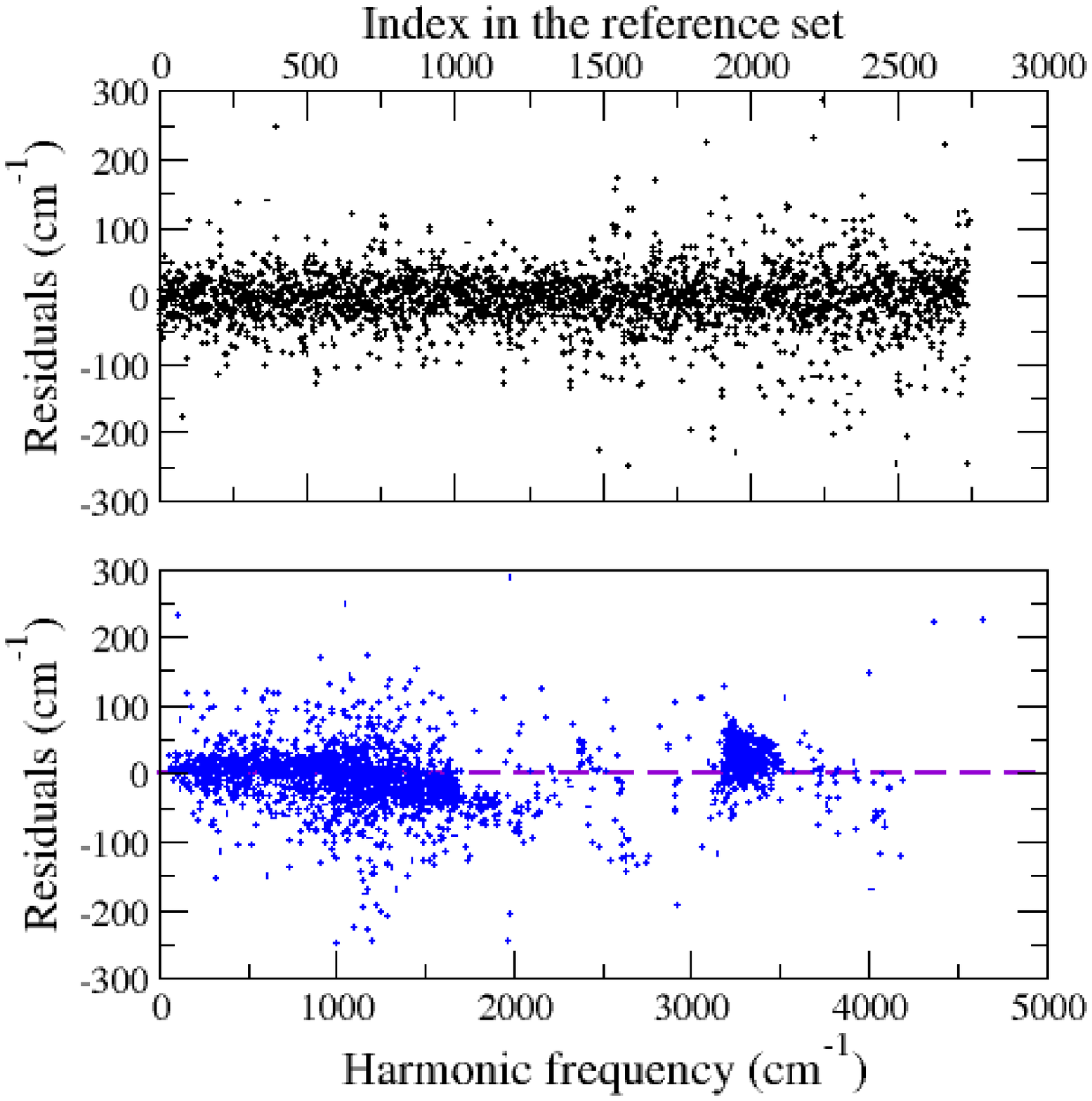}(A)\\
\includegraphics[bb=20bp 30bp 520bp 530bp,clip,width=0.5\columnwidth]{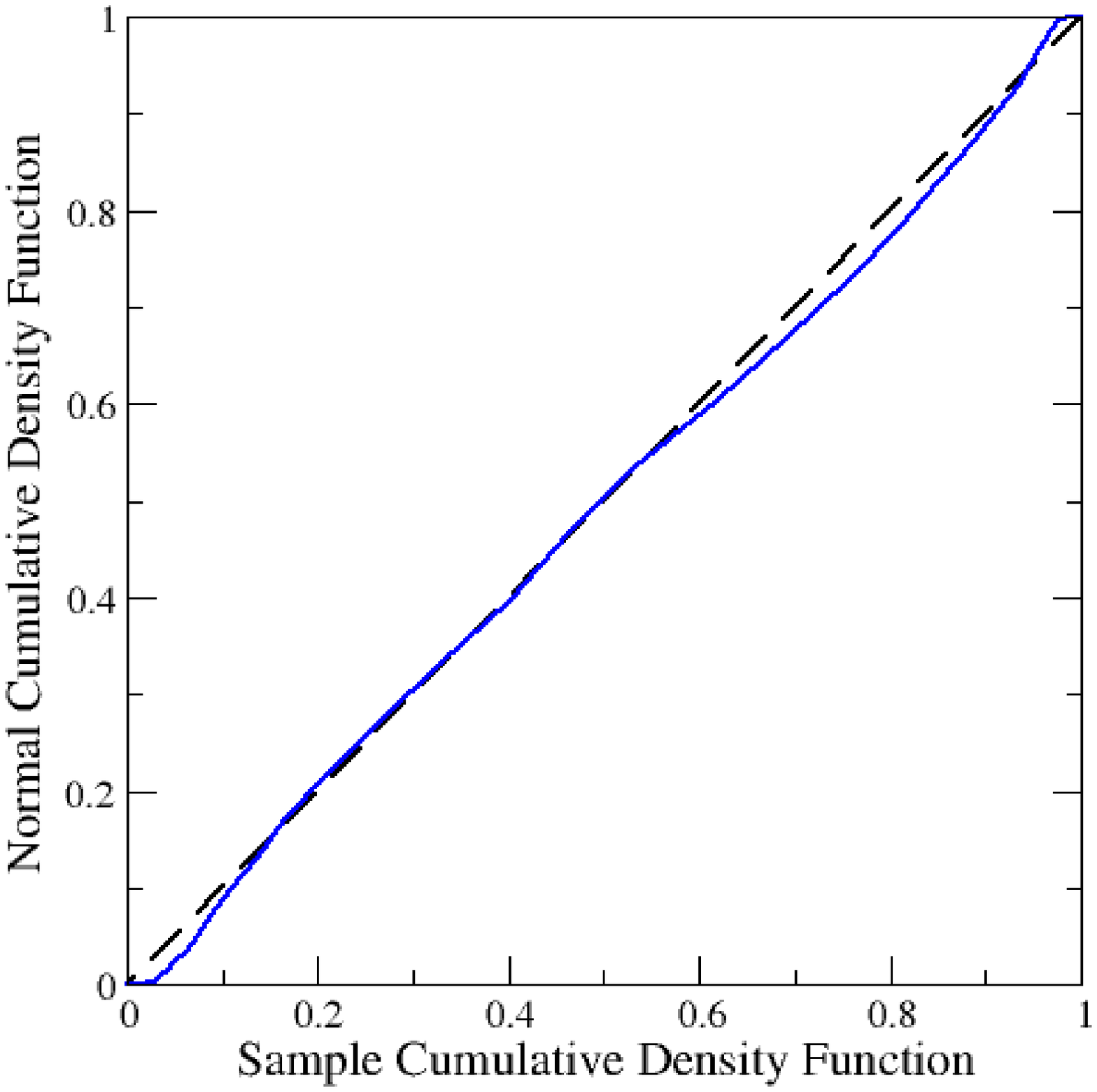}(B)
\par\end{centering}

\caption{\label{fig:Residues}Residuals between calculated harmonic frequencies
$\omega_{i}$ and measured frequencies $\nu_{i}$ for a set of vibrations
extracted from the CCCBDB for the HF/6-31G{*} combination of theory/basis-set
(dots). (A) bottom: residuals as a function of $\omega$. In order
to suppress the grouping effect linked with frequencies, the residuals
were also plotted as a function of their order in the reference set
(A)-top. (B): plot of the cumulative density function (CDF) for the
residuals against a normal CDF shows that globally there is very little
deviation from normality.}

\end{figure}

An option would be to seach for a better model than the linear scaling,
which is beyond the purpose of this study. Instead, we consider here
that the misfit is not deterministically predictable. The solution
to preserve the linear scaling model in a statistically valid model,
is thus to introduce a stochastic variable $\xi$ to represent the
discrepancy between model and observations \begin{equation}
\nu^{obs}=s\omega+\xi+\epsilon\label{eq:calibration model}\end{equation}
This equation is similar to the basic statistical model introduced
by Kennedy and O'Hagan for Bayesian calibration of model outputs \cite{Kennedy01}.
The discrepancy variable could formally depend on $\omega$, but we
observed that the residuals between model and observations are not
notably frequency dependent (Fig. \ref{fig:Residues}), and the same
is assumed for $\xi$ which is considered null in average with variance
$\sigma^{2}$\begin{equation}
\xi\sim N(0,\sigma^{2})\end{equation}
The calibration equation depends thus on two unknown parameters $s$
and $\sigma$. 

\subsection{Uncertainty propagation}

The calibration model (Eq. \ref{eq:calibration model}) being linear
with regard to uncertain variables $s$, $\xi$ and $\epsilon$, and
in the hypothesis of normally distributed uncertainties, we can use
the standard uncertainty propagation rules \cite{GUM} to estimate
the value and uncertainty on predicted vibrational frequencies: \begin{align}
\overline{\nu} & =\overline{s}\,\omega\\
Var(\nu) & =\left(\frac{\partial\nu}{\partial s}\right)_{s=\overline{s}}^{2}Var(s)+\left(\frac{\partial\nu}{\partial\xi}\right)_{\xi=0}^{2}Var(\xi)+\left(\frac{\partial\nu}{\partial\epsilon}\right)_{\epsilon_{}=0}^{2}Var(\epsilon)\\
 & =\omega^{2}Var(s)+\sigma^{2}+\rho^{2}\label{eq:UP_glob}\end{align}
The terms due to covariances have been omitted from this equation
as the three variables are independent by hypothesis. In order to
provide evaluated predictions of vibrational frequencies, we need
therefore to estimate $\overline{s}$, $u_{s}^{2}=Var(s)$ and $\sigma^{2}$
from the calibration dataset. This is done in the next section, using
Bayesian model calibration.

\subsection{Bayesian Model Calibration (BMC)}

\subsubsection{General case}

Starting from the statistical calibration model (Eq. \ref{eq:calibration model}),
one derives the following expression for the posterior probability
density function of the parameters, given a set of $N$ measured and
calculated frequencies (details are reported in Appendix \ref{sub:Taking-measurement-uncertainties})
\begin{eqnarray}
p(s,\sigma|\left\{ \nu_{i}^{obs},\rho_{i},\omega_{i}\right\} _{i=1}^{N}) & \propto & \frac{1}{\sigma\prod_{i=1}^{N}\left(\sigma^{2}+\rho_{i}^{2}\right)^{1/2}}\,\exp\left(-\frac{1}{2}\sum_{i=1}^{N}\frac{\left(\nu_{i}^{obs}-s\omega_{i}\right)^{2}}{\sigma^{2}+\rho_{i}^{2}}\right)\end{eqnarray}
Inferences from this pdf have to be done generally by numerical methods
\cite{Gregory05}, as will be applied later to zero point vibrational
energies. For vibrational frequencies, we first derive an adapted,
simplified, model.

\subsubsection{The case of negligible measurement uncertainties}

In the commonly met situation where the amplitude of $\xi$ is much
larger than the other sources of uncertainty ($\sigma\gg\rho$), we
can consider the approximate measurement model\begin{equation}
\nu_{i}^{obs}=s\omega_{i}+\xi\label{eq:meas_eq}\end{equation}
for which the posterior pdf can be simplified to\begin{eqnarray}
p(s,\sigma|\left\{ \nu_{i}^{obs},\omega_{i}\right\} _{i=1}^{N}) & \propto & \frac{1}{\sigma^{N+1}}\,\exp\left(-\frac{N\gamma^{2}}{2\sigma^{2}}\right)\,\exp\left(-\frac{(s-\overline{s})^{2}\sum_{i=1}^{N}\omega_{i}^{2}}{2\sigma^{2}}\right)\end{eqnarray}
This expression is amenable to analytical derivation of the estimates
for the parameters (see Appendix \ref{sub:Case-of-negligible}):

\begin{itemize}
\item the average value for $s$ is identical to the optimal value provided
by least-squares analysis $\overline{s}=\hat{s}$;
\item the standard uncertainty on $s$, $u_{s}$ is related to the rms $\gamma$
by the formula \\
$u_{s}=\gamma\,\sqrt{N/\left((N-3)\sum\omega_{i}^{2}\right)}$;
\item the estimate of $\sigma^{2}$ is related to $\gamma$ according to
$\overline{\sigma^{2}}=\gamma^{2}\, N/(N-3)$.
\end{itemize}
Using these estimates, we can establish an explicit expression for
the standard uncertainty of $\nu$: \begin{equation}
u_{\nu}=\gamma\,\sqrt{\frac{N}{N-3}\left(\frac{\omega^{2}}{\sum_{i}\omega_{i}^{2}}+1\right)}\label{eq:additive_UP}\end{equation}
Confidence intervals can be defined for prediction purpose, \emph{e.g.
}the 95\% confidence interval for $\nu$, assuming the normality of
uncertainty is given by\begin{equation}
CI_{95}(\nu)=[\hat{s}\omega-1.96\, u_{\nu},\hat{s}\omega+1.96\, u_{\nu}]\end{equation}
It can be seen that for large calibration sets of few hundreds of
frequencies $\sqrt{N/(N-3)}\simeq1$ and $\omega^{2}/\sum_{i}\omega_{i}^{2}\ll1$,
and thus \begin{equation}
u_{\nu}\simeq\gamma\label{eq:UP_simple}\end{equation}
In such conditions, it is thus possible to derive directly confidence
intervals from the the summary calibration statistics $\hat{s}$ and
$\gamma$ provided by the reference literature.\cite{Scott96,Wong96,Merrick07}

\subsection{The Multiplicative Uncertainty (MU) method}

Irikura \emph{et al.} \cite{Irikura05}, after a thorough analysis
of the uncertainty sources, established that the major contribution
to uncertainty propagation would be the uncertainty on the scaling
factor $\hat{s}$. They estimate $u_{s}$ from the weighted variance
of $s$ with weights $a_{i}=\omega_{i}^{2}$. This weighting scheme
is derived in two steps: (1) they propose that the probability density
function (pdf) for the scaling factor is a linear combination of pdf's
for individual scaling factors in the reference set; and (2) from
the comparison of the expression of the average value derived from
this proposition with the least-squares solution Eq.\ref{eq:optim_s}.
This way, they obtain a standard uncertainty\begin{equation}
u_{s}^{*}\simeq\left(\frac{1}{\sum\omega_{i}^{2}}\sum\omega_{i}^{2}\left(s_{i}-\hat{s}\right)^{2}\right)^{1/2}\label{eq:us_irikura}\end{equation}
which can be related to $\gamma$ by $u_{s}^{*}\simeq\gamma\,\sqrt{N/\sum\omega_{i}^{2}}$.
This uncertainty is different, and generally smaller, than the dispersion
of $s$ values within the calibration set \begin{equation}
\sigma_{s}=\left(\frac{1}{N}\sum\left(s_{i}-\hat{s}\right)^{2}\right)^{1/2}\label{eq:sigma_s}\end{equation}
The uncertainty on a scaled frequency is then reduced to \begin{equation}
u_{\nu}\simeq\omega u_{s}^{*}\label{eq:UP_irikura}\end{equation}
hence the name of ''Multiplicative Uncertainty'' (MU) used hereafter.

The salient feature of Eq. \ref{eq:UP_irikura} is that uncertainty
should be proportional to the calculated harmonic frequency. But,
if one observes correlation plots for reference datasets (\emph{e.g.}
Fig. \ref{fig:Correlation-plot}, $\omega>2000$ cm$^{-1}$), this
is not the case. When contrasted with the BMC approach, one understands
that the multiplicative approach ''absorbs'', at least partially,
model inadequacy in $u_{s}^{*}$. It is thus implicitly assumed in
this approach that the model (Eq. \ref{eq:simple model}) is in a
statistical regime, although it is not. These points will be illustrated
and discussed in the next section.

\section{Applications and discussion\label{sec:Applications-and-discussion}}

\subsection{Vibrational frequencies}

We illustrate the BMC and MU approaches on the HF/6-31G{*} combination
of theory/basis-set. The reference dataset, plotted in Fig. \ref{fig:Correlation-plot},
has been downloaded from the NIST/CCCBDB in July 2008.\cite{cccbdb}

\subsubsection{Calibration\label{sub:Calibration}}

\begin{figure}[p]
\noindent \begin{centering}
\includegraphics[bb=0bp 0bp 670bp 612bp,clip,width=1\columnwidth]{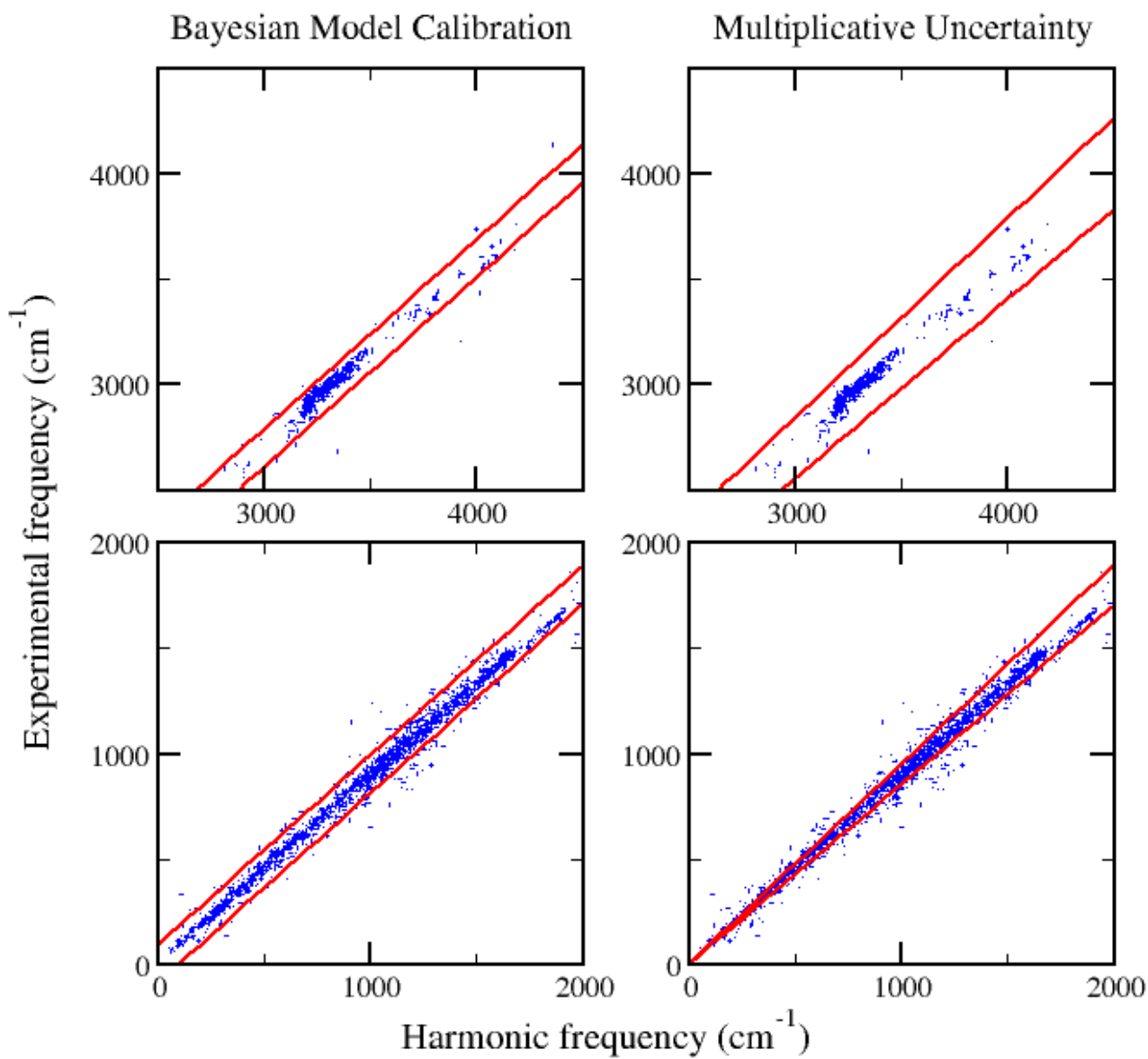}
\par\end{centering}

\caption{\label{fig:CI}Confidence intervals at 95\% level calculated with
Bayesian Model Calibration (left column) and Multiplicative Uncertainty
(right column) methods, in two representative frequency regions from
the calibration dataset for the HF/6-31G{*} combination of theory/basis-set
(dots).}

\end{figure}

Using the BMC model and estimators on the reference set, we obtain
$\hat{s}=0.8984\pm0.0005$, and $\hat{\sigma}=45.3\pm0.6$~cm$^{-1}$
(Table \ref{tab:full_set}), which is consistent with the value of
the rms obtained by Merrick \emph{et al.} \cite{Merrick07} for the
same theory/basis-set combination. For this dataset, the CCCBDB proposes
$\hat{s}=0.899\pm0.025$. We cross-checked this value using Eq. \ref{eq:us_irikura}
(Table \ref{tab:full_set}). The standard uncertainty on $\hat{s}$
evaluated by both methods differ thus by a factor 50, which is expected
to have noticeable effect on prediction uncertainty. In order to visualize
this effect, we plotted the 95 percent confidence intervals in both
cases (Fig. \ref{fig:CI}). It is immediately visible that the the
MU approach has a tendency to underestimate uncertainty at low frequencies
and to overestimate it at high frequencies. In comparison, the BMC
approach is more balanced.

\begin{sidewaystable}
\noindent \begin{centering}
\begin{tabular}{lcccccccccc}
\toprule 
 & \multicolumn{3}{c}{Summary stat.} &  & \multicolumn{2}{c}{MU} &  & \multicolumn{3}{c}{BMC}\tabularnewline
\cmidrule{2-4} \cmidrule{6-7} \cmidrule{9-11} 
 & $\hat{s}$ & $\sigma_{s}$ & $\gamma$(cm$^{-1}$) &  & $u_{s}$ & \%CI$_{95}$ &  & $u_{s}$ & $\sigma$(cm$^{-1}$) & \%CI$_{95}$\tabularnewline
\midrule
\multicolumn{11}{l}{\textbf{All frequencies} (2738)}\tabularnewline
Full set  & 0.8984 & 0.069 & 45.3 &  & 0.024 & - &  & 0.0005 & 45.3 & -\tabularnewline
Calibration set & 0.8986 &  & 45.3 &  & 0.024 & - &  & 0.0007 & 45.3 & -\tabularnewline
Validation set & - & - & - &  & - & 83.0 &  & - & - & 94.6\tabularnewline
\midrule
\multicolumn{11}{l}{\textbf{Frequencies between 3180 and 3500 cm$^{-1}$} (479)}\tabularnewline
Full set  & 0.9050 & 0.0087 & 28.7 &  & 0.0087 & - &  & 0.0004 & 28.8 & -\tabularnewline
Calibration set & 0.9052 &  & 23.3 &  & 0.0071 & - &  & 0.0005 & 23.4 & -\tabularnewline
Validation set & - & - & - &  & - & 95.0 &  & - & - & 95.4\tabularnewline
\bottomrule
\end{tabular}\caption{\label{tab:full_set}Statistical estimates and validation for MU and
BMC models for vibrational frequencies extracted from the CCCBDB for
the HF/6-31G{*} combination of theory/basis-set.}

\par\end{centering}
\end{sidewaystable}

\subsubsection{Validation}

In order to comfort the observations of the previous section (\ref{sub:Calibration}),
we split the calibration dataset in two subsets by using the odd indexes
for calibration, and the even ones for validation (the frequencies
being ordered per molecule, this provides a quasi-random selection
with regard to frequency value). In this case, one gets slightly different
values of the parameters, as reported in Table \ref{tab:full_set}.
Using these values, we generate 95\% confidence intervals and calculate
the frequency of inclusion of the experimental values of the validation
subset within these intervals (Fig. \ref{fig:CI}). For a consistent
estimator, one should find a frequency close to 95\%. BMC succeeds
for 94.7\%, whereas the MU model succeeds for only 83\% of the frequencies
in the validation set (Table \ref{tab:full_set}). Considering the
size of tha samples, the difference is significative, and the consistency
of the multiplicative uncertainty approach in this context can be
questioned.

\textcolor{red}{}

\subsubsection{Test on a restricted frequency scale}

As stated in Paper I, ''to apply the fractional bias correction,
it is important to select a class of frequencies similar to the ones
to be estimated''. For instance, if one selects in the reference
set only those frequencies between 3180 and 3500 cm$^{-1}$, one gets
a much more uniform picture than with the full reference set. The
calibration procedure was done with this limited set of 479 frequencies,
providing $\hat{s}=0.9050\pm0.0087$ (Table \ref{tab:full_set}).
In this case, we note that the uncertainty factor for $s$ is practically
identical to the standard deviation calculated from the sample (0.00869
vs. 0.00871): $u_{s}^{*}\simeq\sigma_{s}$. Due to the restricted
frequency range, one has indeed $\omega_{i}^{2}/\sum_{i}\omega_{i}^{2}\simeq1/N$
(\emph{cf.} Eqs \ref{eq:us_irikura} and \ref{eq:sigma_s}). 

The restricted set has been split in two using index parity, as before.
The scaling factor obtained by MU from the calibration subset is now
$\hat{s}=0.9052\pm0.0071$, and 95.0\% of the validation frequencies
fall within the 95\% confidence interval. This result is now identical
to the one obtained with BMC (Table \ref{tab:full_set}), the confidence
intervals obtained by both methods being indistinguishable.

It appears thus that in restrictive conditions, the MU method can
be valid for reference sets where the individual scaling factors are
uniformly distributed with regard to the harmonic frequencies. In
such case however, the uncertainty is reduced to a conventional unweighted
standard deviation.

\subsubsection{Uncertainty propagation}

The relative importance of both factors $u_{s}^{2}$ and $\sigma^{2}$
($\rho=0$ in this example) in Eq. \ref{eq:UP_glob} can be evaluated
on the example of a calculated harmonic frequency in the higher range
$\omega=3000$ cm$^{-1}$ (Table \ref{tab:UP}). In this case, the
uncertainty on the scaling factor contributes only to one thousandth
of the total prediction variance. 

For any practical purpose, the uncertainty on the scaling factor of
vibrational frequencies can therefore be neglected. The uncertainty
on $\sigma$ is also much too small to be relevant for confidence
intervals calculation. One can thus safely apply the uncertainty propagation
formula (Eq. \ref{eq:UP_simple}), using the rms provided by most
reference articles dealing with scaling factors calibration \cite{Scott96,Wong96,Merrick07}.

\begin{center}
\begin{table}[p]
\noindent \begin{centering}
\begin{tabular}{lllll}
\hline 
 & $\omega$ ({\footnotesize HF/6-31G{*})} & $\omega^{2}u_{s}^{2}$ & $\sigma^{2}$ & $\nu\pm u_{\nu}$\tabularnewline
\hline
Freq. & 3000~ & 2.25 & 2052.09 & 2695$\pm$45~cm$^{-1}$\tabularnewline
ZPVE & 100~kJ~mol$^{-1}$ & 0.029 & 0.194 & 98.12$\pm$0.47~kJ~mol$^{-1}$\tabularnewline
\hline
\end{tabular}\caption{\label{tab:UP}Uncertainty propagation with calibrated model for a
set of 2738 vibrational frequencies extracted from the CCCBDB for
the HF/6-31G{*} combination of theory/basis-set and for a set of 39
ZPVE of the Z1 set for the B3LYP/6-31G{*} combination.}

\par\end{centering}
\end{table}

\par\end{center}

\subsection{Zero Point Vibrational Energies}

We consider ZPVE as a an additional test because the reference datasets
are considerably smaller than for the vibrational frequencies (e.g.
39 molecules in the Z1 set of Merrick \emph{et al.} \cite{Merrick07}),
which is expected to enhance the role of the uncertainty on the scaling
factor. In the absence of a systematic review of measurement errors
for ZPVE of polyatomic molecules, we consider in the following that
they can be neglected. The uncertainties reported by Irikura \cite{Irikura07}
for diatomic molecules are indeed very small (on the order of 0.01
cm$^{-1}$), but transposition to larger molecules is not straightforward.

Using our measurement model and estimators on the reference set, one
gets $\hat{s}=0.9135\pm0.0027$ and $\hat{\sigma}=0.73\pm0.09$~kJ~mol$^{-1}$
(Table \ref{tab:Z1}), which is consistent with the rms obtained by
Merrick \emph{et al.} \cite{Merrick07} for the HF/6-31G{*} theory/basis-set
combination. Relative uncertainties on these parameters have been
increased by one order of magnitude, compared with the vibrational
frequencies case. The validation exercise shows once more that the
Multiplicative Uncertainty model fails to provide correct confidence
intervals, with a score of only 0.63 for CI$_{95}$.

In such a case of small reference dataset, it is interesting to check
if the approximate formula (Eq. \ref{eq:UP_simple}) for uncertainty
propagation which was validated for large sets of vibrational frequencies
begins to break down, \emph{i.e.} the contribution of the multiplicative
term involving $u_{s}$ stays negligible or not, for the larger ZPVE
values. For instance, if one considers a calculated ZPVE of 100 kJ~mol$^{-1}$,
one has $u_{\nu}=\sqrt{(100*0.0027)^{2}+0.73^{2}}=0.78$ kJ~mol$^{-1}$,
to be compared to $\gamma=0.71$ kJ~mol$^{-1}$. It is to be noted
also that the uncertainty on $\sigma$ might contribute at the same
level, $u_{\sigma}=0.09$ kJ~mol$^{-1}$. Taking all uncertainty
sources into account trough Eq. \ref{eq:UP_exact} by Monte Carlo
Uncertainty Propagation, one gets $u_{\nu}=0.77$ kJ~mol$^{-1}$.
Apparently, the fluctuations of $\sigma$ do not have an effect and
only the uncertainty on the scaling factor has a noticeable effect,
albeit quite small.

In the same conditions, for the combination B3LYP/6-31G{*}, one gets
$\gamma=0.423$ kJ~mol$^{-1}$ and $u_{\nu}=0.472$ kJ~mol$^{-1}$,
to be compared with an exact value of $u_{\nu}=0.466$ kJ~mol$^{-1}$
(Table \ref{tab:Z1}). There is globally only a 10\% increase compared
to the rms $\gamma$. In this range of ZPVEs, $\gamma$ still provides
a good approximation of the uncertainty factor (Table \ref{tab:UP}).
However, the amplitude of the discrepancy between $\gamma$ and $u_{\nu}$
will probably increase with the size of the molecule. In consequence,
for uncertainty propagation with ZPVEs, notably for large molecules,
it would be safer to use the full UP formula (Eq. \ref{eq:additive_UP}),
involving the multiplicative uncertainty factor. Compilations of scaling
factors should thus report the easily calculated value of $u_{s}=\gamma\,\sqrt{N/\left((N-3)\sum\omega_{i}^{2}\right)}$,
in addition to the rms $\gamma$.

\begin{sidewaystable}
\noindent \begin{centering}
\begin{tabular}{lcccccccccc}
\toprule 
 & \multicolumn{3}{c}{Summary stat.} &  & \multicolumn{2}{c}{MU} &  & \multicolumn{3}{c}{BMC}\tabularnewline
\cmidrule{2-4} \cmidrule{6-7} \cmidrule{9-11} 
 & $\hat{s}$ & $\sigma_{s}$ & $\gamma$(kJ~mol$^{-1}$) &  & $u_{s}$ & \%CI$_{95}$ &  & $u_{s}$ & $\sigma$(kJ~mol$^{-1}$) & \%CI$_{95}$\tabularnewline
\midrule
\multicolumn{11}{l}{\textbf{HF/6-31G{*} }}\tabularnewline
Full set  & 0.9135 & 0.0607 & 0.71 &  & 0.0161 & - &  & 0.0027 & 0.73$\pm$0.09 & -\tabularnewline
Calibration set & 0.9078 &  & 0.77 &  & 0.0214 &  &  & 0.0052 & 0.83$\pm$0.14 & \tabularnewline
Validation set &  &  &  &  &  & 0.63 &  &  &  & 0.95\tabularnewline
\midrule
\multicolumn{11}{l}{\textbf{B3LYP/6-31G{*}}}\tabularnewline
Full set  & 0.9812 & 0.0375 & 0.42 &  & 0.0103 & - &  & 0.0017 & 0.44$\pm$0.05 & -\tabularnewline
Calibration set & 0.9825 &  & 0.45 &  & 0.0134 &  &  & 0.0032 & 0.48$\pm$0.08 & \tabularnewline
Validation set &  &  &  &  &  & 0.68 &  &  &  & 1.00\tabularnewline
\bottomrule
\end{tabular}\caption{\label{tab:Z1}Statistical estimates and validation for MU and BMC
models for a set of 39 ZPVEs of the Z1 set for the HF/6-31G{*} and
B2LYP/6-31G{*} combinations of theory/basis-set.}

\par\end{centering}
\end{sidewaystable}

\subsubsection{Non-negligible experimental uncertainties}

\begin{figure}[p]
\noindent \begin{centering}
\includegraphics[bb=0bp 0bp 550bp 570bp,clip,width=0.75\columnwidth]{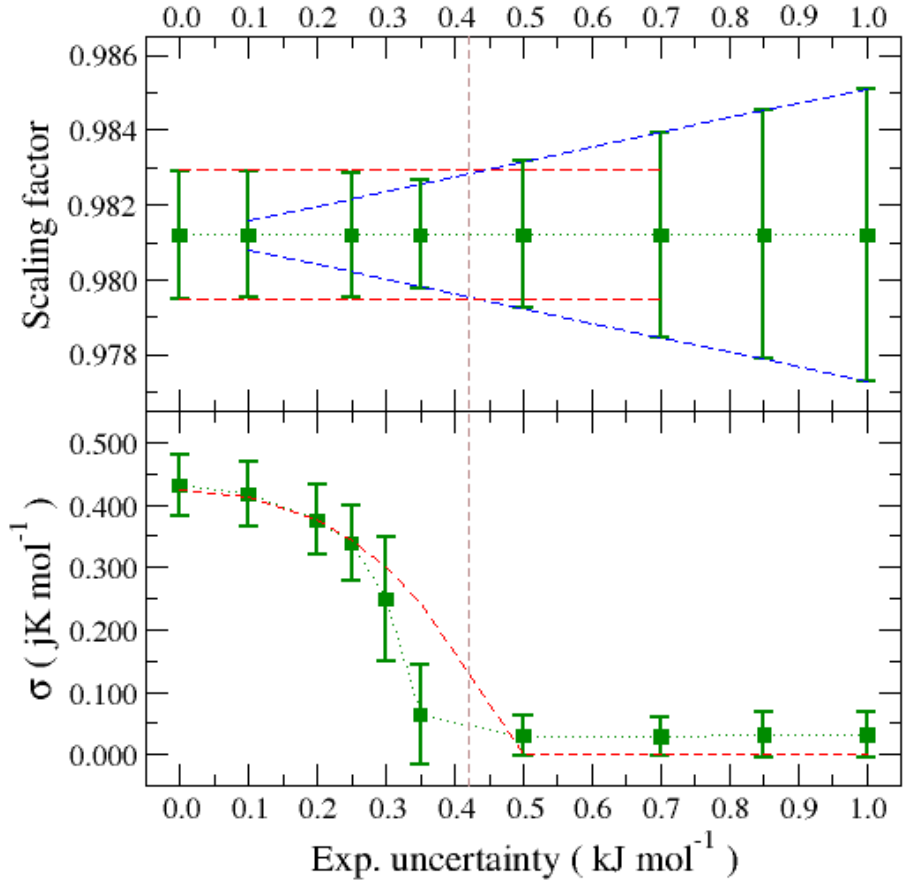}
\par\end{centering}

\caption{\label{fig:Uncert_sc}Evolution of measurement model parameters with
the amplitude of an hypothetical uniform experimental measurement
uncertainty $\rho$ on ZPVE; B3LYP/6-31G{*} combination of theory/basis-set
(green squares with error bars). The brown vertical dashed line indicates
the value of the rms $\gamma$. Top panel: the red dashed lines represent
the 1$\sigma$ confidence interval in the limit of null experimental
uncertainty; the blue dashed line represent the 1$\sigma$ confidence
interval in the weighted least squares limit. Bottom panel: the red
dashed line represents the $\sigma^{2}+\rho^{2}=\gamma^{2}$ law,
truncated to positive values of $\sigma$.}

\end{figure}

When the measurement uncertainty becomes comparable to the rms, model
discrepancy should be vanishing, and confidence intervals for prediction
should include the measurement uncertainty, \emph{i.e.} $u_{\nu}^{2}=\omega^{2}u_{s}^{2}+\sigma^{2}+\rho^{2}$.
In the absence of an exhaustive compilation of experimental uncertainties
on measured ZPVE, we performed simulations assuming a uniform uncertainty
distribution over the full dataset. In order to test the sensitivity
of the model parameters to this uncertainty, we repeated the estimations
of previous section, using Eq. \ref{eq:bayes_exp_unc}, for different
values of $\rho$ between 0.1 and 1.0~kJ~mol$^{-1}$. The results
are reported in Fig. \ref{fig:Uncert_sc}.

As expected from the properties of the posterior pdf, the average/optimal
value of the scaling factor is insensitive to the amplitude of $\rho$.
Moreover, we observe only a slight increase of $u_{s}$ from 0.002
to 0.004. A transition from a constant $u_{s}$, defined by the $\rho=0$
limit, to a linear increase consistent with the weighted least squares
limit (Eq. \ref{eq:wls_lin}) is observed around $\rho=\gamma$, where
both limits intersect. A closer look shows that the transition occurs
indeed at values of $\rho$ slightly smaller than $\gamma$, in a
region ($\rho\simeq0.35)$ where $u_{s}$ displays a minimum. 

The evolution of $\sigma$ is more dramatic: it displays a sharp decrease
and falls down to zero as soon as the measurement uncertainty reaches
and overpasses the value of the rms $\gamma$. For values of $\rho$
below $0.25$, $\sigma$ obeys to the $\sigma^{2}+\rho^{2}=\gamma^{2}$
law (represented as a dashed line in Fig. \ref{fig:Uncert_sc}), but
the calculated decrease becomes much faster in the transition zone.
The uncertainty on $\sigma$ increases notably in the transition region
observed for $s$.

In the limit of large experimental uncertainties, the uncertainty
propagation formula can be written as \begin{align}
u_{\nu}^{2} & =\omega^{2}u_{s}^{2}+\rho^{2}\\
 & =\rho^{2}\left(1+\omega^{2}/\sum\omega_{i}^{2}\right)\end{align}
In this region of large measurement uncertainty, the inadequacy variable
$\xi$ becomes useless, as the calibration with the scaling factor
alone enters a statistical regime.

\section{Conclusion}

A reanalysis of scaling factors calibration in the aim of providing
quantified predictions shows that for large validation sets of accurate
data, as used for vibrational frequencies, faithful prediction uncertainties
can be derived from the standardly reported optimal scaling factor
and rms. For much smaller datasets of a few dozens of data, as in
the case of ZPVEs, uncertainty on the scaling factor should also be
reported. The following limit formulas have been validated and are
proposed for uncertainty propagation from a given calculated harmonic
frequency $\omega$:

\begin{itemize}
\item large calibration sets of precise data ($\rho\ll\gamma$): $u_{\nu}(\omega)=\gamma$;
\item small calibration sets of precise data ($\rho\ll\gamma$): $u_{\nu}(\omega)=\gamma\,\sqrt{\frac{N}{N-3}\left(1+\omega^{2}/\sum_{i}\omega_{i}^{2}\right)}$;
\item sets of data with large uniform uncertainty ($\rho\geq\gamma$): $u_{\nu}(\omega)=\rho\,\sqrt{1+\omega^{2}/\sum_{i}\omega_{i}^{2}}$.
\end{itemize}
A general estimation framework, based on Bayesian Model Calibration,
has been defined for those cases where the limit conditions defined
above are not met.

The Multiplicative Uncertainty method proposed by Irikura \emph{et
al.} \cite{Irikura05} ($u_{\nu}\simeq\omega\gamma\,\sqrt{N/\sum\omega_{i}^{2}}$)
has been shown to be inconsistent when large frequency ranges are
considered. It could not be recovered by the Bayesian approach, except
when the dataset spans a restricted frequency range, in which case
it is reduced to a trivial, unweighted, standard deviation. The use
of the scaling factor uncertainties as reported presently (November
2008) in the CCCBDB\cite{cccbdb} cannot therefore be recommended.

I would like to stress out that the validity of the formulas for uncertainty
propagation depends to some extent on the normality of the residuals
$\nu_{i}-\hat{s}\omega_{i}$ of the linear regression. Inspection
of histograms of residuals (see \emph{e.g.} Fig.1 in Ref. \cite{Wong96})
shows that this is not always the case. The customary approach to
choose an optimal theory/basis-set combination is to assess their
performance by the rms alone, maybe weighted by computational cost
considerations \cite{Scott96,Wong96,Merrick07}. One could also consider
a ''normality criterion'' in order to reject theory/basis-set combinations
providing non-normal residuals and from which the summary statistics
cannot be used faithfully for uncertainty propagation. Analysis of
restricted ranges of data as presently done by some authors for vibrations\cite{Merrick07}
is one way to improve the normality of residuals.

Semi-empirical correction of ab-initio results by scaling is presently
very common and efficient for many observables. It certainly would
be a large step towards the general applicability of uncertainty propagation,
if statistically pertinent estimators were systematically reported
in the literature devoted to the calibration of these correction models.

\section*{Acknowledgments}

I would like to thank Prof. Leo Radom for providing me with the Z1
ZPVE dataset. B. Lévy is warmly acknowledged for helpful discussions.

\appendix

\section{Appendix}

\subsection{Bayesian analysis of scaling factor calibration model\label{sub:Taking-measurement-uncertainties}}

We consider the measurement model \begin{equation}
\nu_{i}^{obs}=s\omega_{i}+\xi+\epsilon_{i}\end{equation}
where $\epsilon_{i}\sim N(0,\rho_{i}^{2})$ is the measurement uncertainty
of $\nu_{i}^{obs}$, and $\xi\sim N(0,\sigma^{2})$ is a discrepancy
function between the linear model and the observations. This model
has two unknown parameters, $s$ and $\sigma$, to be estimated on
a calibration dataset consisting of $N$ calculated harmonic frequencies
$\left\{ \omega_{i}\right\} _{i=1}^{N}$, and their corresponding
observed frequencies $\left\{ \nu_{i}^{obs}\right\} _{i=1}^{N}$. 

In the Bayesian data analysis framework,\cite{Sivia96,Gregory05}
all information about parameters can be obtained from the joint posterior
pdf $p\left(s,\sigma|\left\{ \nu_{i}^{obs},\rho_{i},\omega_{i}\right\} _{i=1}^{N}\right)$.
In order to simplify the notations, we will omit in the following
the list indices when they are not necessary.This pdf is obtained
through Bayes theorem\begin{eqnarray}
p(s,\sigma|\left\{ \nu^{obs},\rho,\omega\right\} ) & \propto & p(\left\{ \nu\right\} |s,\sigma,\left\{ \rho,\omega\right\} )\, p(s,\sigma)\label{eq:Bayes}\end{eqnarray}
where $p(\left\{ \nu^{obs}\right\} |s,\sigma,\left\{ \rho,\omega\right\} )$
is the likelihood function and $p(s,\sigma)$ is the prior pdf. 

As the difference between observation and corrected frequency has
a normal distribution \begin{equation}
\nu_{i}^{obs}-s\omega_{i}\sim N(0,\sigma^{2}+\rho_{i}^{2})\end{equation}
the likelihood function for a single observed frequency is\begin{equation}
p(\nu_{i}^{obs}|s,\sigma,\rho_{i},\omega_{i})=\left(2\pi\left(\sigma^{2}+\rho_{i}^{2}\right)\right)^{-1/2}\exp\left(-\frac{1}{2}\frac{\left(\nu_{i}^{obs}-s\omega_{i}\right)^{2}}{\sigma^{2}+\rho_{i}^{2}}\right)\end{equation}
Considering that all frequencies are measured independently (with
uncorrelated uncertainty) the joint likelihood is the product of the
individual ones, \emph{i.e.}\begin{eqnarray}
p\left(\left\{ \nu^{obs}\right\} |s,\sigma,\left\{ \rho,\omega\right\} \right) & = & \prod_{i=1}^{N}\left(2\pi\left(\sigma^{2}+\rho_{i}^{2}\right)\right)^{-1/2}\exp\left(-\frac{1}{2}\sum_{i=1}^{N}\frac{\left(\nu_{i}^{obs}-s\omega_{i}\right)^{2}}{\sigma^{2}+\rho_{i}^{2}}\right)\end{eqnarray}
As there is a priori no correlation between $s$ and $\sigma$, we
use a factorized prior pdf $p(s,\sigma)=p(s)p(\sigma)$. In the absence
of a priori quantified information on $s$, a uniform pdf $p(s)=cte$
is used. For $\sigma,$we have to consider a positivity constraint,
and we use a Jeffrey's prior, $p(\sigma)\propto\sigma^{-1}$. The
posterior pdf is finally defined up to a proportionality factor which
is irrelevant for the following developments\begin{eqnarray}
p(s,\sigma|\left\{ \nu^{obs},\omega,\rho\right\} ) & \propto & \sigma^{-1}\prod_{i=1}^{N}\left(\sigma^{2}+\rho_{i}^{2}\right)^{-1/2}\exp\left(-\frac{1}{2}\sum_{i=1}^{N}\frac{\left(\nu_{i}^{obs}-s\omega_{i}\right)^{2}}{\sigma^{2}+\rho_{i}^{2}}\right)\label{eq:bayes_exp_unc}\end{eqnarray}

\subsection{Case of negligible measurement uncertainties\label{sub:Case-of-negligible}}

For the analysis of vibrational frequencies, it is generally considered
that experimental uncertainties are negligible ($\rho_{i}=0$). The
general expression for the posterior pdf (Eq. \ref{eq:bayes_exp_unc})
can be simplified accordingly to\begin{eqnarray}
p(s,\sigma|\left\{ \nu^{obs},\omega\right\} ) & \propto & \sigma^{-N-1}\,\exp\left(-\frac{1}{2\sigma^{2}}\sum_{i=1}^{N}\left(\nu_{i}^{obs}-s\omega_{i}\right)^{2}\right)\end{eqnarray}
Using the identity \begin{equation}
\sum_{i=1}^{N}\left(\nu_{i}^{obs}-s\omega_{i}\right)^{2}=(s-\hat{s})^{2}\sum\omega_{i}^{2}+N\gamma^{2}\end{equation}
it can be written in a convenient factorized form\begin{eqnarray}
p(s,\sigma|\left\{ \nu^{obs},\omega\right\} ) & \propto & \sigma^{-N-1}\,\exp\left(-\frac{N\gamma^{2}}{2\sigma^{2}}\right)\,\exp\left(-\frac{(s-\overline{s})^{2}\sum\omega_{i}^{2}}{2\sigma^{2}}\right)\end{eqnarray}
from which we can derive analytical estimates for the parameters and
their uncertainties.

\subsubsection{Estimation of $s$}

The marginal density for $s$ is obtained by integration over $\sigma$

\begin{eqnarray}
p(s|\left\{ \nu^{obs},\omega\right\} ) & = & \int_{0}^{\infty}d\sigma\, p(s,\sigma|\left\{ \nu^{obs},\omega\right\} )\\
 & \propto & \int_{0}^{\infty}d\sigma\,\sigma^{-N-1}\exp\left(-\frac{1}{2\sigma^{2}}\sum_{i=1}^{N}\left(\nu_{i}^{obs}-s\omega_{i}\right)^{2}\right)\\
 & \propto & \left(\sum_{i=1}^{N}\left(\nu_{i}^{obs}-s\omega_{i}\right)^{2}\right)^{-N/2}\label{eq:like_1}\end{eqnarray}
which can be rewritten as \begin{equation}
p(s|\left\{ \nu^{obs},\omega\right\} )\propto\left(1+\frac{(s-\hat{s})^{2}\sum\omega_{i}^{2}}{N\gamma^{2}}\right)^{-N/2}\end{equation}
which has the shape of a Student's distribution \cite{Evans00}

\begin{equation}
\mathrm{Stt}(x)\propto\left(1+\frac{x^{2}}{n}\right)^{-(n+1)/2}\end{equation}
Posing $n=N-1$ and $x^{2}=(N-1)/N\,(s-\hat{s})^{2}\sum\omega_{i}^{2}/\gamma^{2}$,
we can directly use the properties of the Student's distribution \begin{equation}
\mathrm{E}[x]=0;\,\mathrm{Var}[x]=n/(n-2)\end{equation}
to derive\begin{align}
\overline{s} & =\hat{s}\\
u_{s}^{2} & =\gamma^{2}\,\frac{N}{(N-3)\sum\omega_{i}^{2}}\end{align}

\subsubsection{Estimation of $\sigma$}

The marginal density for the standard uncertainty of the stochastic
variable $\xi$ is \begin{eqnarray}
p(\sigma|\left\{ \nu^{obs},\omega\right\} ) & = & \int_{-\infty}^{\infty}ds\, p(s,\sigma|\left\{ \nu^{obs},\omega\right\} )\\
 & \propto & \frac{1}{\sigma^{N+1}}\exp\left(-\frac{N\gamma^{2}}{2\sigma^{2}}\right)\int_{-\infty}^{\infty}ds\exp\left(-\frac{(s-\hat{s})^{2}\sum\omega_{i}^{2}}{2\sigma^{2}}\right)\\
 & \propto & \frac{1}{\sigma^{N}}\exp\left(-\frac{N\gamma^{2}}{2\sigma^{2}}\right)\end{eqnarray}
Using the formula\begin{equation}
\int_{0}^{\infty}dx\, x^{-n}e^{-a/x^{2}}=\frac{1}{2}\Gamma\left(\frac{n-1}{2}\right)/a^{(n-1)/2}\end{equation}
one obtains readily the following estimates\begin{align}
\hat{\sigma} & =\gamma\\
\overline{\sigma} & =\gamma\,\sqrt{\frac{N}{2}}\frac{\Gamma\left[(n-2)/2\right]}{\Gamma\left[(n-1)/2\right]}\\
\overline{\sigma^{2}} & =\frac{N}{N-3}\gamma^{2}\\
u_{\sigma} & =\gamma\,\sqrt{\frac{N}{N-3}-\frac{N}{2}\left(\frac{\Gamma\left[(n-2)/2\right]}{\Gamma\left[(n-1)/2\right]}\right)^{2}}\end{align}

\subsection{Case of very large measurement uncertainties\label{sub:Case-of-very}}

When model discrepancy is negligible compared to measurement uncertainties,
\emph{i.e.} when the standard linear model is in a statistical regime,
one recovers standard statistical analysis, the Bayesian analog to
weighted least squares. The posterior pdf for $s$ is then\begin{eqnarray}
p(s|\left\{ \nu^{obs},\omega,\rho\right\} ) & \propto & \prod_{i=1}^{N}\rho_{i}^{-1}\exp\left(-\frac{1}{2}\sum_{i=1}^{N}\frac{\left(\nu_{i}^{obs}-s\omega_{i}\right)^{2}}{\rho_{i}^{2}}\right)\end{eqnarray}
from which one obtains\begin{align}
\hat{s} & =\sum\left(\omega_{i}\nu_{i}^{obs}/\rho_{i}^{2}\right)/\sum\left(\omega_{i}^{2}/\rho_{i}^{2}\right)\\
u_{s}^{2} & =1/\sum\left(\omega_{i}^{2}/\rho_{i}^{2}\right)\end{align}
For uniform experimental uncertainty over the dataset, the scaling
factor uncertainty varies linearly with $\rho$ \begin{equation}
u_{s}=\rho/\sqrt{\sum\omega_{i}^{2}}\label{eq:wls_lin}\end{equation}

\subsection{Uncertainty propagation}

In the Bayesian framework, the posterior pdf $p(s,\sigma|\left\{ \nu^{obs},\rho,\omega\right\} )$
can be used to estimate the uncertainty of predicted frequencies\begin{equation}
p(\nu'|\rho',\omega',\left\{ \nu^{obs},\rho,\omega\right\} )=\int dsd\sigma\, p(\nu'|s,\sigma,\rho',\omega')p(s,\sigma|\left\{ \nu^{obs},\rho,\omega\right\} )\label{eq:UP_exact}\end{equation}
where\begin{equation}
p(\nu'|s,\sigma,\rho',\omega')\propto(\sigma^{2}+\rho'^{2})^{-1/2}\exp\left(-\frac{\left(\nu'-s\omega'\right)^{2}}{2(\sigma^{2}+\rho'^{2})}\right)\end{equation}
results from the stochastic model (Eq. \ref{eq:meas_eq}). This integral
has generally to be evaluated numerically.

\newpage{}\bibliographystyle{unsrt}

\end{document}